\documentclass[pre,twocolumn,showpacs,amsmath,amssymb]{revtex4}

\usepackage{amsmath}
\usepackage{amssymb}
\usepackage{graphicx}
\usepackage{color}
\usepackage{bm}
\newcommand{\B}[1]{{\bm{#1}}}

\definecolor{g-blue}{rgb}{0.83,0.95,1}
\definecolor{g-yellow}{rgb}{1,1,0.7}
\definecolor{g-green}{rgb}{0.9,1,0.9}
\definecolor{green}{rgb}{0,0.6,0}
\definecolor{cyan}{rgb}{0,0.7,0.7}
\definecolor{black}{rgb}{0,0,0}
\definecolor{grey}{rgb}{0.4 ,0.4 ,0.4 }

\def\blue#1{\textcolor{blue}{#1}}
\def\red#1{\textcolor{red}{#1}}
\def\green#1{\textcolor{green}{#1}}

\begin{document}
\title{Kelvin-wave turbulence in superfluids}
\date{\today}
\author{Laurent Bou\'e, Ratul Dasgupta, Jason Laurie$^{\dagger}$, Victor L'vov, Sergey Nazarenko$^\ddagger$ and Itamar Procaccia}

\affiliation{Department of Chemical Physics, Weizmann Institute of Science, Rehovot 76100, Israel \\
$^\dagger$~Laboratoire de Physique, ENS Lyon, 46 All\'{e}e d'Italie, F69007 Lyon, France \\
$^\ddagger$~Mathematics Institute, University of Warwick, Coventry CV4 7AL, United Kingdom
}

\begin{abstract}
We study the statistical and dynamical behavior of turbulent Kelvin waves propagating on quantized vortices in superfluids,  and address the  controversy concerning the energy spectrum that is associated with these excitations.  Finding the correct energy spectrum is important because Kelvin waves play a major role in the dissipation of energy in superfluid turbulence at near-zero temperatures. In this paper, we show analytically that the solution proposed in Ref.~\cite{10LN} enjoys existence, uniqueness and regularity of the pre-factor.  Furthermore, we present numerical results of the dynamical equation that describes to leading order
 the non-local regime of the Kelvin wave dynamics.  We compare our findings with the analytical results from the proposed local and non-local theories for Kelvin wave dynamics and show an agreement with the non-local predictions.
 Accordingly, the spectrum proposed in Ref.~\cite{10LN} should be used in future theories of quantum turbulence.
Finally, for weaker wave forcing we observe an intermittent behavior of the wave spectrum with a fluctuating dissipative scale, which we interpreted as a finite-size effect characteristic to mesoscopic wave turbulence.
\end{abstract}
\maketitle

\section{\label{s:intro}Introduction}

When sufficiently large waves propagate through a medium, non-linear interactions cause an energy transfer between different scales;  the energy that is introduced typically in a narrow band of wavelength is redistributed to waves of different wavelengths.  Starting in the 1960s with oceanographers and meteorologists, `weak wave turbulence' has now become a standard theory to explain the dynamical and statistical properties of an ensemble of weakly nonlinear interacting waves~\cite{92ZLF}.  Thanks to the generality of its formulation, this theory is an interdisciplinary subject; it has been implemented in situations ranging from Alfv\'en waves in solar winds, ocean swells on a stormy sea, quantum waves in Bose-Einstein condensates, Rossby waves in the atmospheres of rotating planets as well as in the thunder-like sound produced by vibrating thin elastic sheets, etc.  Up-to-date discussions of the current state of wave turbulence are available for example in a recent book ~\cite{11Nazar} and in reviews~\cite{10Fal,11NR}; these sources also contain extensive lists of references.

Quantum turbulence is the study of turbulent behavior in zero temperature superfluids such as helium II~\cite{D91, 02VN,KS09}.  One of the key properties of quantum turbulence is that it comprises of an inviscid flow which permits energy to reach scales far smaller than achieved in classical turbulence.  At large scales, quantum turbulence shows a great similarity to classical turbulence - with the polarization of vortex bundles acting like large scale eddies.  However, at small scales the analogy begins to break down.  In the absence of viscosity in superfluids and at very low temperatures where thermal excitations (mutual friction) can be neglected, there is no clear dissipation mechanism that cuts off the energy cascade to very small lengthscales.  Therefore, unlike in classical turbulence where viscous effects dissipate the energy, it is believed that for superfluids the energy is transferred to propagating waves on quantized vortex lines that are forced by vortex reconnections.  Note that these so-called ``Kelvin waves'' were originally introduced in~1880 in the context of the classical Rankine vortex model~\cite{80Tho}. The Kelvin waves then interact allowing for energy to be transferred to higher frequency Kelvin waves until the wave frequency is sufficient for the excitation of phonons and thus the degradation of energy into heat~\cite{D91}.

Because of its importance in superfluid turbulence and the growing experimental capabilities in this field~\cite{02VN,08Nie,09TH}, there has recently been a renewed interest in the statistical physics of waves propagating on a vortex line~\cite{01KVSB,03VTM,04KS, 10LLNR}.  A complete understanding of the statistical behavior of Kelvin waves is therefore crucial in order to develop a theory of superfluid turbulence.  Until recently, Kelvin wave interactions were thought to be a local process - only Kelvin waves of similar wave numbers interact with each other thus forming the range of a local energy cascade~\cite{04KS}.  The Kelvin wave cascade was described using wave turbulence  theory~\cite{92ZLF,11Nazar} for weakly interacting waves, resulting in the formation of a kinetic wave equation describing   $3 \leftrightarrow 3$ Kelvin wave interactions~\cite{04KS}. However in Ref.~\cite{10LLNR}, the locality assumption used in Ref.~\cite{04KS} was checked and was shown to be violated for local $3 \leftrightarrow 3$ Kelvin wave interactions.  This invalidated the local theory, and a non-local theory was proposed~\cite{10LN} resulting in $1 \leftrightarrow 3$ Kelvin wave interactions.  This has prompted a lively debate about  the correct spectrum of Kelvin waves~\cite{LL10,KS10,LLN10,KS10-1,KS10-2}.

\subsection{\label{ss:physics} Physical background}

It is now recognized that the typical turbulent state of a superfluid consists of a complex tangle of quantized vortex lines~\cite{93Don} which can be modelled by the Biot-Savart equation~\cite{D91}
\begin{equation}\label{Biot-Savart equation}
 \frac{d   \bm  r}{d   t} = \frac{\kappa}{4\pi}\int \frac{({\bm  s}-{\bm  r}) \times d{\bm  s}}{|{\bm  r}-{\bm  s}|^3}\ .
\end{equation}
This equation contains a singularity as ${\bm  s} \to {\bm  r}$, and hence the model are postulated by introducing a cut-off in the integration: $a_0 < |{\bm  r}-{\bm  s}|$, where $a_0$ is interpreted as the core radius of the vortices \cite{S85,S88}.  Here, $\kappa=2\pi \hbar/M$ is the quantum of circulation with $M$ being $m_4$, mass of   $^4$He atom (if the superfluid is  $^3$He then $M$ is the mass of two atoms, 2$m_3$).

An important step in studying Kelvin wave turbulence was done by Sonin~\cite{S85} and later by Svistunov~\cite{S88} who found a Hamiltonian form of the Biot-Savart Eq.~(\ref{Biot-Savart equation})  for a straight vortex line aligned in the $z$ direction, e.g. line $(x,y)=(00)$. Perturbing the line by small disturbances in the $(x,y)$-plane,
 \begin{subequations}\label{eq:hamsys}
 \begin{equation}\label{eq:hamsysA}
 w(z,t) = x(z,t)+iy(z,t),
 \end{equation}
 one writes:
 \begin{equation}\label{eq:hamsysB}
 i\kappa\frac{\partial w}{\partial t} = \frac{\delta H }{\delta w^*},
\end{equation}\end{subequations}
 where $\delta(\dots)/\delta w^*$ is the functional derivative of $(\dots)$ and the superscript ``$\,^*\, $" denotes complex conjugation. The Hamiltonian for the Biot-Savart equation $H$ is the  energy of the system \cite{87Son,95Svi}:
\begin{equation}\label{Ham}
 H = \frac{\kappa^2}{4\pi} \int \frac{1+\Re(w'^*(z_1)w'(z_2))}{\sqrt{(z_1-z_2)^2+|w(z_1)-w(z_2)|^2}}  dz_1 dz_2\ .
\end{equation}
Here we have used the notation $w'(z)=dw/dz$.

Like in most of the examples of wave turbulence mentioned in the introduction, the theory for Kelvin waves starts by writing down a Hamiltonian equation for the complex canonical wave amplitudes~$a_{\bm k} (t)$ and~$a_{\bm k}^* (t)$ which are  classical analogs of the creation and annihilation Bose operators in quantum mechanics \cite{87Son,95Svi}.

Consider an isolated straight vortex line on
a periodic domain of length $ {\cal  L} $.
One can write \cite{92ZLF,11Nazar}
\begin{equation}
i\,  \frac{d a_{\bm k}(t)}{d t} = \frac{\partial \mathcal{H}}{\partial a_{\bm k}^*(t)}\ ,
\end{equation}
where $w(z,t) = \kappa^{-1/2} \sum_{\bm k} a_{\bm k}(t) \exp (i\B k\cdot \B z )$.  The new Hamiltonian  $ {\cal   H}\{a ,a ^*\}\equiv  H\{w,w^* \} /  {\cal  L} $  is the  density of the old one and is a function of all $a_{\B k}(t)$ and  $a_{\B k}^*(t)$ taken at the same time.

For small Kelvin wave amplitudes (inclination angles) the Hamiltonian can be expanded
 with respect of  $ \green{a_{\B k} \,, a_{\B k}^*}$. The explicit form of $H$, Eq.~(\ref{Ham}), dictates an expansion of  $\cal H$   in $a_{\B k}$ and $a_{\B k}^*$ with  even powers only,
  \begin{subequations}
   \begin{equation} \label{exp-H}
   {\cal  H } = {\cal  H}_2+  {\cal  H}\sb{\rm int}\,, \quad {\cal  H}\sb{\rm int}={\cal  H}_4+ {\cal  H}_6+\dots
 \end{equation}
  The first term
 \begin{equation} \label{H2}
  \green{ {\cal H}_2  = \sum_{\B k}  \omega_k\, a_{\B k}\, a^*_{\B k}\,,}
     \end{equation}
      describes free propagation of Kelvin waves with a frequency   $ \omega_k $. In  turn, $\omega_k$ should be expanded in inverse powers of the large parameter $\Lambda$:
        \begin{eqnarray}\nonumber
  \omega_k
  &\simeq&  ^ \Lambda \omega_k + ^ 1 \omega _k\,,
   \   ^ \Lambda \omega_k =\frac{\kappa\,  \Lambda}{4\pi} \,{k^2}\,,\
    ^1\omega_k =  - \frac{\kappa\, \ln \big(k\ell\big)}{4\pi}\, k^2  \,, \\ \label{omega}
    \Lambda &=& \ln(\ell /a)\ .
    \end{eqnarray}
  Here   $a$ is the vortex line diameter and~$\ell$ is the mean inter-vortex distance at which the description of Kelvin waves propagating along an individual vortex line fails. In typical experiments $\Lambda$, in both  $^3$He and  $^4$He, is between 12 and 15 \cite{09TH}. It can be shown~\cite{04KS} that the leading approximation in $\Lambda$ gives no energy exchange between Kelvin waves  and
therefore one has to account in $\cal H$ for subleading terms, zero order in $\Lambda$, denoted by the superscript ``$^1$".

The higher order expansion terms in $  {\cal  H}_{\rm int}$, $  {\cal  H}_4 $  and    ${ \cal  H}_6$,  describe  $2 \leftrightarrow 2$  and $3 \leftrightarrow 3$ scattering of Kelvin waves:
 \begin{eqnarray}\label{Ham4}
{\cal  H}_4&=& \frac 14 \sum_{1+2=3+4} T_{\,1,2}^{3,4}\ a_1a_2a_3^* a_4^* \,,  \quad a_j\equiv a(\B k_j,t)\,,~~~\\
   {    \cal  H}_6 &=& \frac1{36} \sum_{1+2+3=4+5+6}\!\!\!  W_{\,1,2,3}^{4,5,6}\ a_1a_2a_3 a_4^*a_5^* a_6^*\,,
\end{eqnarray}
\end{subequations}
Equations for the  terms of order  $\Lambda^1$ and  $\Lambda^0$ in the interaction amplitudes $T_{\,1,2}^{3,4} \equiv  T(\B k_1,\B k_2|\B k_3,\B k_4)$ and  $W_{\,1,2,3}^{4,5,5} \equiv  W(\B k_1,\B k_2,\B k_3|\B k_4,\B k_5,\B k_6)$  (denoted as $  ^\Lambda T_{\,1,2}^{3,4}, \   ^1 T_{\,1,2}^{3,4}$ and  $  ^ \Lambda W_{\,1,2,3}^{4,5,6}, \ ^ 1 W_{\,1,2,3}^{4,5,6} $ )
were found in Ref.~\cite{04KS} and later confirmed in
 Ref.~\cite{10LLNR}.

On the face of it, the leading term in $\mathcal{H}_{\mbox{\tiny int}}$ (i.e. ${\cal H}_4$) describes a $(2\leftrightarrow 2)$ scattering; the subleasing term, ${\cal H}_6$, is  responsible for the $(3\leftrightarrow 3)$ scattering.
It was argued, however, that the $(2\leftrightarrow 2)$  scattering cannot redistribute
 energy between different scales. Therefore  the $(3\leftrightarrow 3)$ scattering becomes the leading interaction  responsible for the inter-scale energy transfers.  This  does not mean, however, that the ${\cal H}_4$ Hamiltonian can be completely disregarded. Instead, as explained in Refs.~\cite{92ZLF,10LLNR},  ineffective Hamiltonian (in our case ${\cal H}_4$) can be eliminated from the problem by a proper non-linear canonical transformation  $\{ a, a^*\} \Rightarrow \{ b, b^*\}$.  This comes at a price of the appearance of ``correction terms" in the 6-wave mixing  amplitude (which we have denoted with a calligraphic vertex $\cal W$, different from $W$) in  the Hamiltonian  $\widetilde{{\cal H}}_6$:
\begin{subequations} \label{Heff}
\begin{eqnarray}\label{tildeH}
\widetilde{\cal  H}&=&  \green{\sum_{\B k}  \omega_k\, b_{\B k}\, b^*_{ \B k}}+ \widetilde{{\cal H}}_6\,, \\ \widetilde{{\cal H}}_6&=&
\frac1{36} \sum_{1+2+3=4+5+6}\!\!\!   {\cal W}_{\,1,2,3}^{4,5,6}\ b_1b_2b_3 b_4^*b_5^* b_6^* \ .
 \end{eqnarray}
 Note that the 4-wave mixing Hamiltonian has been eliminated altogether by the canonical transformation $\{ a, a^*\} \Rightarrow \{ b, b^*\}$.
 The remnant of this Hamiltonian  appears as an additional contribution to $\cal W$.  There is an exact relation between these two vertices which is best represented by a schematic graphic notation as
 \begin{eqnarray}\nonumber
&&\includegraphics[width=0.8\linewidth]{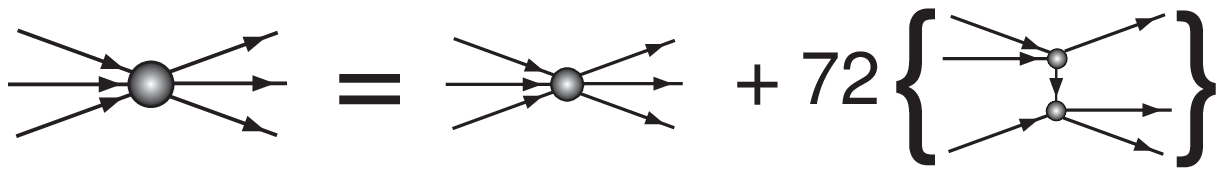},\\ \label{graph}
 &&~~ {\cal W}_{\,1,2,3}^{4,5,6}~~~~~= ~~~  W _{\,1,2,3}^{4,5,6}~~~~+ 72~~ \, \{ ~~T^2 / \, \Omega ~~\}\ .~~~~~~~~
\end{eqnarray}\end{subequations}
The 72 additional contributions are schematically indicated as $72\{...\}$ since the exact expression is too long to be written here (one can handle them with a symbolic computation software such as {\em Mathematica}). One observes that, on the one hand, they are 6-wave mixing terms, but due
to their internal structure they can also be understood as the pairs of 4-wave mixing amplitudes (mediated by a Green's function $1/\Omega$). In the graph shown in \eqref{graph}, with incoming wave vectors  ${\bm k}_1, \ {\bm k}_2,\  {\bm k}_3$ starting from above  and outgoing wave vectors ${\bm k}_4, \ {\bm k}_5, \ {\bm k}_6$,  the frequency $\Omega=\omega({\bm k}_1)+\omega({\bm k}_2)-\omega({\bm k}_4)-\omega({\bm k}_1+{\bm k}_2-{\bm k}_4)$.
The 72 contributions of the type shown in \eqref{graph} differ by the directions of the arrows and by relabeling in ${\bm k}_1, \ {\bm k}_2,\  {\bm k}_3$ and ${\bm k}_4, \ {\bm k}_5, \ {\bm k}_6$ groups.

This understanding provides us
with an effective 6-wave mixing, allowing one to use standard procedure~\cite{92ZLF,11Nazar} for
the statistical description of weak turbulence of Kelvin waves. It is based on the assumption that the turbulent dynamics  of waves with small enough amplitude is
chaotic and creates its own ergodic measure. The simplifying nature of weak wave turbulence
is that, due to the existence of a small
parameter, the statistical description closes upon itself in terms of the pair correlation function
\begin{subequations} \label{KE}
\begin{equation}\label{def-n}
n_{\B k} (t) \equiv \frac{\cal  L} {2\pi} \green{\langle b_{\B k}(t) b^*_{\B k}(t)\rangle} \,,
\end{equation}
which is also called the ``wave action". Hereafter
 the pointed brackets stand for an average over the ergodic measure.  The aim of the theory is to
analyze the solutions of the kinetic equation  of motion which is typically expressed as
\begin{equation}
\frac{d n _{\bm k}(t)}{d t} = \mbox{St}(\B k,\{n_{\B k'}(t)\})\,,
\label{kinetic}
\end{equation}\end{subequations}
where  term $\mbox{St}(\B k,\{n_{\B k'}(t)\})$ is so-called
collision integral with an integrand proportional to the square of the effective    interaction amplitude $\cal W$. This term is a function of $\B k$ and a functional of~$n_{\B k'}(t)$~\cite{92ZLF,11NR}.

The main part of the Kelvin wave energy $E$ in the regime of weak wave turbulence is given by $E_2= \langle  {\cal H}_2 \rangle$, where ${\cal H}_2$ is defined   by Eq.~\eqref{H2}. Together with Eq.~\eqref{def-n} this gives
\begin{subequations} \label{KWen}
 \begin{eqnarray}\label{KWen-a}
E&=& \green{\frac{2\pi}{\cal L}}\sum_{\B k}\omega_k n_{\B k} =\green{\int\limits _{-\infty}^{\infty} \omega_k n_{\B k} d  {\B k}} \\  \nonumber &=& \green{\int\limits _0 ^{\infty} \omega_k N_k d  k \equiv \int\limits _0 ^{\infty}   E_k \, d  k\ .}
\end{eqnarray}
Here  we introduce the energy density in the $k$-space,
 \begin{equation}\label{KWen-b}
 E(k)\equiv E_k= \green{\, \omega_k N_k} \,,
  \end{equation}which traditionally is called ``energy spectrum" \green{and we define the ``wave action spectrum" as $N_k = n_{\B k} + n_{-\B k}$, where $k= |{\B k}|$.}
According to Eq.~\eqref{omega} in the leading in $\Lambda$ approximation the energy spectrum of Kelvin waves is related to the wave action $N_k$ as follows:
\begin{equation}\label{KWen-c}
E_k= \frac {\Lambda\,  \kappa}{\green{4}\, \pi}\, k^2\,  N_k\ .
 \end{equation}
\end{subequations}
Up to this point these considerations are agreed by one and all, and are the basis of further developments.

\subsection{\label{ss:KS-LN}The controversy}

Recently Kosik and Svistunov ~\cite{04KS} derived an energy spectrum of Kelvin waves turbulence:
\begin{subequations} \label{KWspectra}
\begin{equation} \label{KS-spectrum}
E_{_{\rm KS}}(k)= C_{_{\rm KS}}\frac{\Lambda\, \kappa^{7/5}\, \epsilon^{1/5}}{ k^{7/5}}\,,
\end{equation}
where $\epsilon$ is the energy flux over scales and $\green{C_{_{\rm KS}}}$ is yet unknown dimensionless constant. Later L'vov and Nazarenko
derived a very different result for the same spectrum:
\begin{equation}\label{LN-spectrum}
E_{_{\rm LN}}(k)= C_{_{\rm LN}}\frac{\Lambda\, \kappa \,  \epsilon^{1/3}}{\Psi^{2/3} \, k^{5/3}}\,,   \quad \Psi \equiv   \frac{\green{\bm 8}\,  \pi \, E}{\Lambda\, \kappa^2}\ .
  \end{equation}
\end{subequations}
Here $C_{_{\rm LN}}$ is another dimensionless constant.
Both spectra are supposed to be ``universal", (i.e. independent of details of the energy forcing) in the the scaling range $k > k\sb{\rm f} $,  where $k_f$ is the forcing wave number.

The disagreement between the spectra  (\ref{KWspectra}) resulted  in a heated debate concerning the correct nature of the energy spectrum.
To identify  the origin of this  controversy  we should clearly state that both results~(\ref{KWspectra}) were obtained within the same formal setup described in the previous subsection under the same set of assumptions
about small nonlinearity and random phases. Therefore the difference between the spectra must originate from one or more mistakes made by either or both derivations. Indeed, we will argue that the mistake leading to the wrong result \eqref{KS-spectrum} is in the
\blue{wrong assumption about } the asymptotic behavior of the effective amplitude
${\cal W}_{\,1,2,3}^{4,5,6}$
 in the region where one of the wave vectors in the interacting sextet is much smaller than at least one other wave vector from the same sextet. The form of this asymptotics crucially affects the nature of the energy transfer in the Kelvin wave cascade.

The derivation of the interaction vertex ${\cal W}_{\,1,2,3}^{4,5,6}$ for Kelvin wave turbulence is not an easy task.  In Ref. ~\cite{04KS} the explicit form of ${\cal W}_{\,1,2,3}^{4,5,6}$ was not presented. Instead, the authors ``simulated the collision integral by a  Monte Carlo method" with the conclusion that it converges and the main contribution to the energy evolution of Kelvin wave with given $k$-wave vector originates from the energy exchange with other Kelvin wave with $k'\sim k$.   This statement of \emph{the locality of the energy transfer} allowed the authors to use a dimensional estimate that leads to the the spectrum (\ref{KS-spectrum}).

This result was criticized in \cite{10LLNR} where  an explicit expression
for the interaction amplitude  ${\cal W}_{\,1,2,3}^{4,5,6}$  (in the asymptotic region) was derived:
 \begin{subequations}\label{W6}
 \begin{equation}\label{LN-W6}
\mathcal{W}^{3,4,5}_{{\bm k},1,2} = - \frac{3 \B k \B k_1 \B k_2 \B k_3 \B k_4 \B k_5}{4\pi \kappa}\ .
\end{equation}

Based on this equation,  it was shown analytically in Ref.~\cite{10LLNR}
that the collision integral diverges. Moreover, as found   in Ref.~\cite{10LLNR}, two important sets of terms in $\mathcal{W}^{3,4,5}_{{\bm k},1,2}$ of the order of unity were overlooked in Ref.~\cite{04KS}. One was the consequence of a trivial algebraic mistake in the Taylor expansion in Eq.~(\ref{W6}), where the authors forgot to expand the denominator. In addition,  in  Ref.~\cite{04KS} there was a  conceptual mistake: in the kinetic equation (see below) the conservation of energy requires the exact frequency-resonance condition that accounts for the sub-leading  contribution $^1\omega_k$,~(\ref{omega}), while cancellation of linear in $\Lambda$ terms in $\mathcal{W}^{3,4,5}_{{\bm k},1,2}$  takes place only on the local induction approximation manifold.  Remaining contributions of the order of unity to $\mathcal{W}^{3,4,5}_{{\bm k},1,2}$ where omitted in  Ref.~\cite{04KS}. It remains unclear, how Ref.~\cite{04KS} succeeded to state convergence of the collision integral; this must be either due to a mistakes in the calculation of $\mathcal{W}^{3,4,5}_{{\bm k},1,2}$ or due to an inaccurate implementation of the Monte Carlo numerical procedure.

The divergence of the collision integral  makes the assumption of locality for the KS spectrum invalid~\cite{10LLNR}.  The non-locality of the spectrum~\eqref{KS-spectrum} implies that it is un-realizable and physically irrelevant.
Recognizing the need for a greater understanding of the correct interaction term, a new {\it non-local} theory was promptly put forward, suggesting the alternative  energy spectrum~(\ref{LN-spectrum}) of Kelvin waves.

This result was immediately attacked in Ref.~\cite{KS10-a}, based on some na\"ive  symmetry
arguments claiming  that
\begin{equation}\label{KS-W6}
\mathcal{W}^{3,4,5}_{{\bm k},1,2} \propto k^2  \,
\end{equation}\end{subequations}
for small $k$ instead of the linear asymptotics~(\ref{LN-W6}).
Note that in the case of quadratic asymptotics~(\ref{KS-W6}) the collision integral converges. Then the Kelvin wave energy cascade would be dominated by local interactions and spectrum~(\ref{KS-spectrum}) would be valid. In the case of linear asymptotics~(\ref{LN-W6}) the collision integral diverges~\cite{10LLNR}, the Kelvin wave energy cascade is dominated by different-scale interactions and the spectrum~(\ref{LN-spectrum}) takes place~\cite{10LN}.

 To conclude,  the controversy regarding the Kelvin wave energy spectra~(\ref{KWspectra}) is a direct consequence of the disagreement regarding the asymptotic behavior of  $\mathcal{W}^{3,4,5}_{{\bm k},1,2}$ for small $k$. This is a fortunate situation, since it can be
 easily resolved by a careful calculation. To this aim, an explicit analytical expression~(\ref{LN-W6}) with its line-by-line derivation was made publicly available ~\cite{11-LBL}. The symmetry arguments of Ref.~\cite{KS10-a} were analyzed in Ref.~\cite{LL10} where it was shown that the symmetry argument appeals to a local reference system in which $z$-axis follows the vortex line curved by long Kelvin waves, while the Hamiltonian and the interaction amplitude must be written in the global reference system with the $z$-axis oriented along  the straight (unperturbed) vortex line. In the other words, the symmetry arguments are irrelevant to the problem under consideration~\cite{LL10}, and see also ~\cite{11Zak} and ~\cite{LLN10}.

We thus feel confident to proceed further in studying Kelvin wave turbulence in the framework of the Hamiltonian~(\ref{Heff}) with the interaction amplitude~(\ref{LN-W6}). The controversy can be reopened only when someone will present a checkable
 derivation of an alternative expression for $\mathcal{W}^{3,4,5}_{{\bm k},1,2}$, which would be different  from Eq.~\eqref{LN-W6}.

\subsection{Some caveats and warnings}

The reader of this paper should not conclude that the problem  of Kelvin waves in superfluids is solved. Rather, this is a complicated and intriguing phenomenon and a lot of actual problems in this field are still open or hardly studied. First, we stress that Kelvin waves in real superfluid turbulence propagate along dynamically bent vortex lines  in a vortex tangle. Although  in the limit of short Kelvin waves the time dependence of the basic vortex configuration and their local curvature can be neglected,
longer waves are also important  in real experimental situations. Secondly and more importantly, even short waves could travel long distances, whereas the classical wave turbulence theory assumes that the wave  "mean free path" is much less that the system size (the classical-quantum crossover in our case). How important such a finite-size is not yet understood.  Numerical simulation of the full Biot-Savart equation which could clarify these issues is still in its infancy~\cite{11BB}.

In this paper, our analytical theory will assume that these  limits of  short and short-correlated Kelvin waves are achieved. In addition we will assume that the amplitudes of Kelvin waves are small. Then the Hamiltonian formulation of the problem~(\ref{Heff},\ref{LN-W6}) is justified. Nevertheless it  does not mean that spectrum~(\ref{LN-spectrum}) is the only possible solution for Kelvin wave turbulence. It known that approximations of weak wave turbulence leading from the {\em dynamical} Hamiltonian equations to the {\em kinetic} equations in one dimensional media are very delicate. Therefore in the second part of this paper we will present direct numerical simulations of Kelvin wave turbulence in the Hamiltonian formulation ~(\ref{Heff},\ref{LN-W6}). This simulation allows us to clarify another important issue: how long should the vortex line be in comparison to the Kelvin wave-length to achieve the limit of unlimited (in the physical space) medium used in the theoretical analysis below. Also, it allows to study the case when this limit is not achieved and the finite-size effects are important - this is the case of so-called mesoscopic wave turbulence.

In Ref.~\cite{10LN} it was  assumed  that the kinetic equation in its continuous-media limit is applicable. The analysis of its collision integral~\cite{10LN,10LLNR} revealed that {\em on the spectrum
\eqref{KS-spectrum}} the leading contribution appears from those terms in which two wave-vectors are much smaller that the other four and a
conjecture was made (not a proof!) that the same kind of nonlocality should hold {\em generally} for most other Kelvin wave spectra, in both steady and evolving turbulent states.  This allowed for the development of an {\em effective} collision integral dominated by four wave mixing of the $(1\leftrightarrow 3)$ type resulting in a new four-wave kinetic equation.  The remaining two small wave-vectors describe the chaotic bending of the vortex lines with a characteristic curvature radius of the order of the inter-vortex distance.  We should note however that even in the framework of the effective four-wave kinetic equations~\cite{10LN} some questions have remained  open. Indeed,  the spectrum (\ref{LN-spectrum}) was proven to be
a valid solution of this equation but its uniqueness in the class of scale invariant solutions and its stability have remained unclear and the pre-factor $C_{_{\rm LN}}$ has not been found. In Section \ref{s:anal}  we will demonstrate that the scale invariant energy spectrum (\ref{LN-spectrum}) exists, it is unique with a regular pre-factor,   $\green{C_{_{\rm LN}}\approx 0.304}$.
Moreover, our numerics of the dynamical equation show that the (\ref{LN-spectrum}) is indeed an attracting solution for the Kelvin wave systems. Not only we observe the correct value of the exponent for this spectrum, but also a relatively good agreement for the pre-factor. This is remarkable because this is probably the first example in wave turbulence where such agreement for the pre-factor value has been reported in numerical simulations. Moreover, these numerical results demonstrate that the type of nonlocality conjectured for the Kelvin waves (two small wave vectors in a typical interacting sextet) is robust and maintained in the attracting steady state.

In the next section we will prove that the non-local theory is correct and is the one that should be used in future theories of quantum turbulence.   After this we will present numerical simulations that are based on the interaction term~(\ref{LN-W6}) and which not only confirm our analytical calculations but also validate the applicability of the kinetic equation. These numerics also allow to study the regimes which are hard to treat analytically when the kinetic equation description breaks down due to the finite-size effects -
so-called mesoscopic wave turbulence.

\section{\label{s:anal} Energy spectrum of Kelvin-wave weak turbulence}

In this section we calculate an exact solution to the new {\it effective} kinetic equation. In~\cite{10LN} it was shown that the collision integral takes the form
\begin{widetext}
\begin{eqnarray}
& & \!\!\!\!\!\!\!\!\!\!\!\!\!\!\!\! \mbox{St}(\B k,\{n_{\B k'}(t)\}) \!= \!\frac{\pi}{12} \iiint\!\! \mbox{d}{\bm k}_1 \mbox{d}{\bm k}_2 \mbox{d}{\bm k}_3 \left( \mathcal{W}_{{\bm k}}^{1,2,3} \right)^2 \, \green{n_{\bm k}} n_1 n_2 n_3 \left[\left( \frac{1}{n_{\bm k}} \!-\! \frac{1}{n_1} \!- \!\frac{1}{n_2} \!-\! \frac{1}{n_3} \right) \delta \left( {\bm k} - {\bm k}_1 - {\bm k}_2 - {\bm k}_3  \right) \delta \left( \omega_k - \omega_1 - \omega_2 - \omega_3  \right) \nonumber \right. \\
& - & 3 \left.\left( \frac{1}{n_1} - \frac{1}{n_{\bm k}} - \frac{1}{n_2} - \frac{1}{n_3} \right) \delta \left( {\bm k}_1\! - \!{\bm k}\! - \!{\bm k}_2 \!- \!{\bm k}_3  \right) \delta \left( \omega_1 - \omega_k - \omega_2 - \omega_3  \right)\right] \ , \quad n_j\equiv \green{n_{{\bm k}_j}} \ ,
\label{collision}
\label{St}
\end{eqnarray}
\end{widetext}
where the {\it effective} $(1\leftrightarrow 3)$ interaction amplitude is
\begin{equation}
\mathcal{W}_{{\bm k}}^{1,2,3} \equiv \frac {-3\Psi}{4\pi\sqrt{2}}\B k\B k_1 \B k_2 \B k_3\ .
\end{equation}
The dimensionless parameter $\Psi$ was defined in~Eq. (\ref{LN-spectrum}).

One obvious steady-state solution of Eq.~(\ref{kinetic}) is $n_k\propto (\omega_k)^{-1}$ which corresponds to the equilibrium Rayleigh-Jeans equipartition of the energy. A second, non-equilibrium solution, can be obtained by analyzing the equation of
energy conservation,
\begin{equation}
\green{\frac{\partial E_{k}}{\partial t} + \frac{\partial \epsilon_{ k}}{\partial { k}} = 0}\ ,
\end{equation}
the energy flux~$\epsilon_{\bm k}$ is given by \footnote{$\epsilon_{\bm k}$ denotes the full energy flux, which in the weakly nonlinear limit is well approximated by the energy flux of the quadratic part of the energy, defined by the right-hand side of Eq.~\eqref{energyeq}. When considering the kinetic equation this definition will be assumed to be the same.}
\begin{equation}
\green{\epsilon_{ k} = -  \int_{{ k'}<{ k}} \mbox{St}_{{\bm k'}} \, \omega_{ {k'}} d{{ k'}}}\ .
\label{energyeq}
\end{equation}
In Ref. \cite{10LN} it was found that the kinetic equation has a scale-invariant solution
\begin{equation}
\green{N_{ k} = A k^{-x}} \ . \label{nofk}
\end{equation}
The scaling exponent $x=11/3$ was found by power counting in the second Eq.~(\ref{energyeq}) under the assumption
that all the integral in the collision term converge. Substituting this solution into the collision integral it
was demonstrated in Ref. \cite{10LN} that the integral indeed converges. Finally the suggested solution has the form \cite{10LN}
\begin{equation}
\green{N_{ k} = C_{_{\rm LN}}\frac{{4} \pi \epsilon^{1/3}}{\Psi^{2/3} \,  k^{11/3}}} \ , \label{sol}
\end{equation}
where the dimensionless coefficient $C_{_{\rm LN}}$ remained undetermined.

Let us now demonstrate analytically that Eq.~(\ref{sol}) is indeed a solution of the kinetic equation. Moreover,
we demonstrate numerically that this solution is the unique scale-invariant solution with a constant energy flux, and
finally we compute the numerical value of the universal constant $C_{_{\rm LN}}$. The first task can be achieved by considering
the second term in Eq. (\ref{St}) as a sum of three equivalent terms. In the first we make the Zakharov-Kraichnan conformal
transformation \cite{92ZLF,11Nazar} from $k_1,k_2,k_3$ to $k'_1,k'_2,k'_3$ where
\begin{equation}
k'_1 = \frac{k^2}{k_1}, \quad k'_2 = \frac{k k_2}{k_1}, \quad k'_3 = \frac{k k_3}{k_1}\ .
\end{equation}
Relabeling then the dummy variable $k'_j\to k_j$ we find that this term, up to a factor $(k_1/k)^\xi$ with
$\xi=3x-9$, coincides with the first term in Eq. (\ref{St}). In the second of the three terms, we perform the
same transformation replacing $k_1 \leftrightarrow k_2$, and in the third term we do the same but replacing
$k_1 \leftrightarrow k_3$. Therefore, the entire collision term in Eq. (\ref{St}) can be represented as the first
term multiplied by
\begin{equation}
[1-(k_1/k)^\xi-(k_2/k)^\xi-(k_3/k)^\xi] \ .
\end{equation}
Note that  the region of integration becomes the same for all the terms too. Due to the existence of the delta-function
of frequencies we conclude that the integrand in the collision term vanishes if $\xi=2$. This condition is
equivalent to $x=x_0=11/3$.

Next we demonstrate the uniqueness of this solution. To do so numerically it is advantageous to non-dimensionalize the
collision integral for arbitrary values of $x$ as follows,
\begin{equation}
\mbox{St}_k = \frac{3 \Psi^2 A^3}{128 \pi \alpha} \, I(x) \, k^{8-3x}\ ,
\label{stk}
\end{equation}
where~$\alpha = \kappa \Lambda / 4 \pi$. Using $q_j\equiv k_j/k$ we obtain
\begin{widetext}
\begin{equation}
I(x) = \iiint \mbox{d}{\bm q}_1 \mbox{d}{\bm q}_2 \mbox{d}{\bm q}_3 \left( 1 - q_1^{\xi} - q_2^{\xi} - q_3^{\xi} \right) \left( 1 - q_1^x - q_2^x - q_3^x \right) \left( q_1 q_2 q_3 \right)^{2-x} \delta \left( 1 - {\bm q}_1 - {\bm q}_2 - {\bm q}_3  \right) \delta \left( 1 - q_1^2 - q_2^2 - q_3^2 \right) \nonumber  \ .\\
\label{adim}
\end{equation}
\end{widetext}

\begin{figure}
\begin{center}
\includegraphics[width=0.8\linewidth]{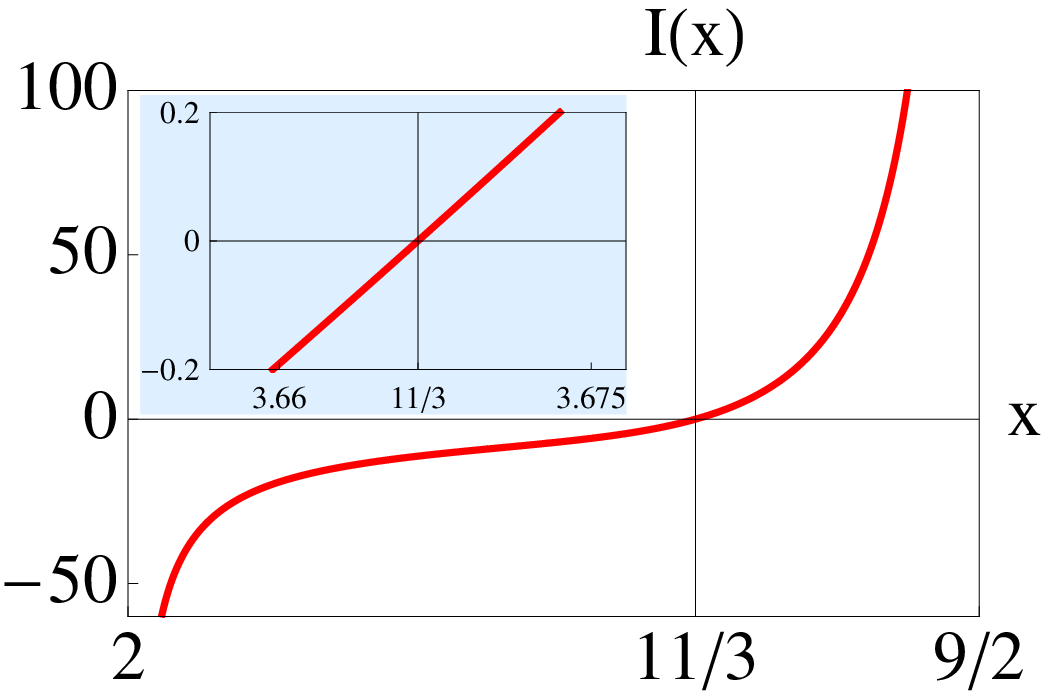}
\caption{The collision integral as a function of the scaling exponent. We observe that it vanishes at~$x=11/3$ and that this cancellation is unique within the window of locality. The insert is just a blow-up of the neighborhood  of $x=11/3$.
One clearly sees that the solution is unique. }
\label{plot}
\end{center}
\end{figure}

This integral was computed numerically as a function of $x$, and plotted in Fig. \ref{plot}, for $2< x < 9/2$ where this
integral converges.
We see that throughout the window of locality (where the integral converges) there is no other solution.

Finally, to compute the universal coefficient $C_{\green{B}}$ we return to Eq.~(\ref{energyeq}) with $\green{N_k}$ given by Eq. (\ref{nofk})
with an arbitrary value of $x$.
Integrating with respect to $\B k'$ leads to an explicit expression for the dissipation rate
\begin{equation}
\epsilon_k = \frac{3\Psi^2 A^3}{\green{128}\pi}\frac{I(x)}{3x-11} k^{11-3x} \ .
\end{equation}
Using L'Hopital rule to deal with the indeterminate ratio of zero by zero we can rewrite this
expression for $x=11/3$,
\begin{equation}
\epsilon_k= \frac{\Psi^2 A^3}{\green{128}\pi}\frac{dI(x)}{dx}\Big|_{x=11/3} \ .
\end{equation}
Computing numerically $dI(x)/dx$ we have an explicit result for the coefficient A leading to
\begin{equation}
\green{N_{ k} =  \frac{4 \pi C_{_{\rm LN}}\epsilon^{1/3}}{\Psi^{2/3} \, k^{11/3}}}; \quad C_{_{\rm LN}}\approx \red{0.304}.
\label{final}
\end{equation}

\noindent Combining this result with the dispersion relation, we finally obtain the energy spectrum
\begin{equation}
E_k = C_{_{\rm LN}}\frac{\Lambda \, \kappa \, \varepsilon^{1/3}}{\Psi^{2/3} \, k^{5/3}}\ .
\label{finalspect}
\end{equation}

\noindent The calculation of the energy spectrum for non-local Kelvin wave interactions constitutes the central result of this section.  We point out that the exponent with~$5/3 \approx 1.67$ is quite close to that of the spectrum~\eqref{KS-spectrum} with~$7/5 \approx 1.4$.  This has made it difficult for numerical simulations to clearly resolve the difference between both theories and as a result has contributed to adding more fuel to the controversy~\cite{03VTM, KS05}.  In fact, a recent numerical simulation using a scale separation scheme for the~Biot-Savart equation produced results consistent with the spectrum~\eqref{KS-spectrum}~\cite{KS10-2}.  However, we believe that the scale separation scheme - which expands the Biot-Savart equation in terms of non-local contributions -  artificially damps the {\em essential} non-local behavior of the Biot-Savart equation, thus inducing local Kelvin wave interaction and the KS spectrum.

\section{\label{s:DNS} Direct numerical simulation of the dynamical equation for the Kelvin waves}

\subsection{\label{ss:LNE}
Dynamical model: Local Nonlinear Equation}

In this section, we present a new simple model that contains all the necessary physics of the~Biot-Savart equation without its numerical complexity.  As mentioned above, an exact analytical expansion of the Hamiltonian~(\ref{Ham}) was carried out in~\cite{10LLNR} leading the authors to the interaction term presented in~Eq.~(\ref{LN-W6}).  This term is the leading order asymptotic of the~Biot-Savart interaction in the limit when at least one Kelvin wave  is much longer than the shortest wave  in each interacting sextet. Obviously, the dynamically important nonlocal sextets in which two waves are long are also described by this asymptotic expression.
When expressed in the physical space, the interaction term~(\ref{LN-W6}) leads to the Local Nonlinear equation~(LNE)~\cite{10LLNR} which is given by
\begin{equation}\label{eq:LNE}
 i\frac{\partial w}{\partial t} = -\frac{\kappa}{4\pi}\frac{\partial}{\partial z}\left[\left(\Lambda - \frac{1}{4}\left|\frac{\partial w}{\partial z}\right|^4\right)\frac{\partial w}{\partial z}\ \right] \, .
\end{equation}
The LNE can be represented as a Hamiltonian system of the form \eqref{eq:hamsys}, with Hamiltonian which in terms of the physical space amplitude  $w(z,t) = x(z,t)+iy(z,t)$ is:
\begin{equation}
 \mathcal{H}^{^{\rm LNE}} = \mathcal{H}_2 +\mathcal{H}_6 =\frac {\kappa^2}{4\pi} \int \left( \Lambda \left |\frac{\partial w}{\partial z}\right |^2 - \frac{1}{12} \left |\frac{\partial w}{\partial z} \right |^6 \right) d z\, .
\end{equation}

\subsection{\label{ss:DNSgamma}
Nonlinear dissipation and applicability of weak turbulence}

To appreciate the conditions under which we could expect realization of the weak wave turbulence regime in our numerics, we should estimate the degree of the nonlinear dissipation given by the nonlinear resonance broadening parameter $\Gamma_k$.
Indeed, the kinetic equation  for weak wave turbulence could only be valid when the nonlinear resonance broadening $\Gamma_k$ is much less than the linear frequency $\omega_k$.
On the other hand, $\Gamma_k$ itself could be estimated from the kinetic equation's collision term \eqref{St}, namely from its part proportional to $n_k$. Importantly, even though the full collision integral converges on the LN spectrum, its separate parts diverge~\cite{10LN}. In particular, the part corresponding to $\Gamma_k$ will scale as
(see Eqs.(19) of Ref.~\cite{10LN})) $\Gamma_k n_k \propto n_k k$ giving
\begin{equation}\label{eq:gamma}
\Gamma_k \propto k\ .
\end{equation}
Thus for the nonlinearity parameter we get
\begin{equation}\label{eq:ratioT}
\frac{\Gamma_k}{\omega_k} \propto k^{-1}\ .
\end{equation}
Eq.~\eqref{eq:ratioT} states that the nonlinearity parameter decreases toward large $k$, i.e. that wave turbulence becomes weaker as one propagates along the energy cascade to high wave numbers.

On the other hand, for validity of the the kinetic equation it is also necessary that the nonlinear resonance broadening $\Gamma_k$ is much greater than the grid spacing between the wave frequencies  $\Delta \omega$, i.e. $\Delta \omega \ll \Gamma_k $.  If the wave turbulence becomes too weak, we are in danger of loosing wave resonances due to the sparsity (discreteness) of the allowed wave numbers and frequencies in bounded domains. If this happens, we expect regimes of discrete or mesoscopic wave turbulence \cite{N06,meso,LNP06,ZKPD05,DLN07,MHDenslaving,11Nazar}.  Mesoscopic wave turbulence occurs when the nonlinear resonance broadening $\Gamma_k$ becomes of the order of the grid spacing between frequencies $\Delta \omega$, i.e. $\Gamma_k \sim \Delta \omega$.  In such a situation, discreteness effects become apparent and the wave amplitudes become too weak to sustain a continuous cascade.  Instead, we observe sandpile behaviour \cite{N06}, where energy accumulates at specific scales until it is reaches sufficient amplitudes to cascade.  This process is then repeated at smaller and smaller scales, until energy eventually reaches the dissipation scale.  When $ \Gamma_k \ll\Delta \omega  $, then we are in a discrete wave turbulence regime, where only exact wave resonances occur.  Resonant waves form  clusters, some of which disjoint from the rest of the modes, and the cascade scenario is even more inhibited.

The frequency grid spacing can be determined via $\Delta \omega =
(\partial \omega_k/\partial k)  \Delta k = (\Lambda \kappa  k /  2 \pi) \Delta k$, where $\Delta k = 2\pi / \mathcal{L}$ is the wave number grid spacing.   Taking into account \eqref{eq:gamma}, one can see that the ratio of $\Gamma_k/\Delta \omega$ is predicted to be independent of $k$
implying that the wave turbulence description should break down when the wave modes are too weak simultaneously and uniformly in the whole of the inertial range of $k$.
\subsection{\label{ss:DNSsetup} Numerical setup}

We perform numerical simulations of the LNE to verify the power law scaling and coefficient of the energy spectrum for Kelvin wave turbulence in a statistically steady state corresponding to an energy cascade from long to short wave scales.  To achieve this, we numerically solve the non-dimensionalized version of Eq.~\eqref{eq:LNE} in the presence of forcing and dissipation:
\begin{equation}\label{eq:nondimLNE}
 i\frac{\partial \tilde{w}}{\partial \tilde{t}} = -\frac{\partial}{\partial \tilde{z}}\left[\left(1 - \frac{1}{4}\left|\frac{\partial \tilde{w}}{\partial \tilde{z}}\right|^4\right)\frac{\partial \tilde{w}}{\partial \tilde{z}}\right] + iF_{\B k} - iD_{\B k}\ ,
\end{equation}
with additive forcing, $+iF_{\B k}$, and dissipation $-iD_{\B k}$.  The non-dimensionalization we use is
 \begin{equation}
 w = \Lambda^{1/4}\tilde{w}, \; t = \frac{4\pi}{\kappa\Lambda}\tilde{t}, \; z = \tilde{z}, \; \tilde{w} = \sum_{\bm k} \tilde{w}_{\bm k}(t) \exp({i\tilde{\B k }\tilde{ z}})\ .
 \end{equation}

We implement a pseudo-spectral method with $2^{14}$ spatial points, where  the nonlinear term is fully de-aliased with respect to its quintic nonlinearity. We utilize a fourth order Runge-Kutta time stepping method to progress in time.  The forcing profile we use is defined in wave number space, as is given by
\begin{equation}
F_{\B k} =
\begin{cases}
 A_f\exp(i\theta_{\B k}) & \text{if } 8\leq k<16\\
0 & \text{otherwise}\ ,
\end{cases}
\end{equation}
where $A_f$ is a constant forcing amplitude and $\theta_{\B k}$ is a random variable sampled from a uniform distribution on $[0,2\pi)$ for each ${\B k}$ and at each time step.  We add dissipation at both ends of the wave number space to enable the formation of a statistically steady state and to avoid any bottleneck effects.  We use a hyper-viscosity term that dissipates mainly at high wave numbers and a low wavenumber friction on the first six wave modes. The dissipation profile $D_{\B k}$ is defined as
\begin{eqnarray}
 D_{\B k} = \nu_{\mathrm{fric}} \; H(6-|{\B k}|)\; \tilde{w}_{\B k} + \nu_{\mathrm{hyper}}\; k^4\; \tilde{w}_{\B k}\ ,
\end{eqnarray}
where,  $\nu_{\mathrm{fric}}=2.0$ and $\nu_{\mathrm{hyper}}=2 \times 10^{-6}$ are dissipation coefficients and $H(\cdot)$ is a Heaviside function.

We perform three simulations, {\bf A}, {\bf B} and {\bf C}, which  have different forcing amplitudes (with the rest of the parameters being identical).  The forcing amplitudes of the simulations are given in Table~\ref{tab:para}.
We allow the simulation to evolve until we have reached a   statistically steady state, judged  by observing stationarity of the total energy, whereupon we average over a time window to achieve the desired statistics.

\begin{table}
\begin{center}
\begin{tabular}{  | c | c | c | c | }
  \hline
   & Sim {\bf A} & Sim {\bf B} & Sim {\bf C}\\
   \hline
   ${A}_f $ & $610$ & $488$ &  $0.6103$\\
  $\tilde{A} $ & $3.2$ & $2.0$ & $0.23$\\
  $\tilde{\epsilon}_k$ & $4.6$ & $4.8$ & $0.55$ \\
  $\tilde{\Psi}$ & $0.35$ & $0.25$ & $0.13$\\
  $|\tilde{\mathcal{H}}_6|/\tilde{\mathcal{H}}_2$ & $0.25$ & $0.20$ & $0.003 - 0.15$\\
  $C_{_{\rm LN}}$ & $0.3477$ & $0.2024$ & $0.031$\\
  \hline
\end{tabular}
\end{center}
\caption{We present the non-dimensional values of the amplitude of the wave action spectrum $\tilde{B}$, energy flux $\tilde{\epsilon}_k$, $\tilde{\Psi}$, the ratio of the nonlinear and linear energies $|\mathcal{H}_6| / \mathcal{H}_2$ and the coefficient  $C_{_{\rm LN}}$ for all three simulations. \label{tab:para}}
\end{table}

\subsection{\label{ss:DNSresults} Numerical results}

\begin{figure}
 \begin{center}
  \includegraphics[width=\columnwidth]{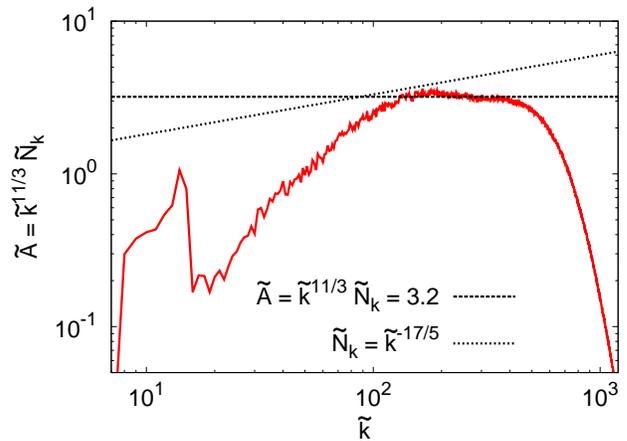}
\caption{Averaged wave action spectrum $\tilde{N}_k$ in simulation {\bf A}, compensated by $k^{11/3}$.    We overlay the theoretical predictions of the Kozik-Svistunov and Lvov-Nazarenko spectra.}
\label{fig:Spectrum}
 \end{center}
\end{figure}

For simulation {\bf A}, we show in Fig.~\ref{fig:Spectrum} the compensated (by $k^{11/3}$) wave action spectrum, and compare against both the predictions of Eq.~\eqref{KS-spectrum} (or, in terms of the wave action, $N_k \sim k^{-17/5}$)  and
Eq.~\eqref{KS-spectrum} (corresponding to $N_k \sim k^{-11/3}$, see~\eqref{finalspect}). The spectrum has been averaged over a long time window once a   statistically steady state is reached. We observe a remarkable agreement with spectrum
suggested by Lvov and Nazarenko, \eqref{sol}, for about a decade in wave number space.  Moreover,  we clearly observe a deviation from the slope of Kozik-Svistunov spectrum, $N_k \sim k^{-11/3}$.

\begin{figure}
 \begin{center}
  \includegraphics[width=\columnwidth]{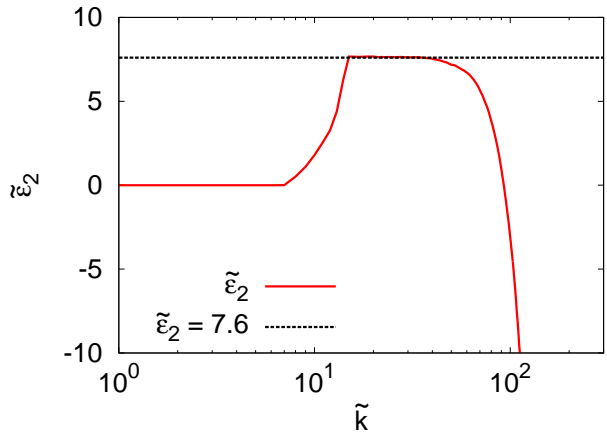}
\caption{Averaged  non-dimensionalized flux of the quadratic energy $\tilde{\epsilon}_2(k)$  in simulation {\bf A}.  We observe a  region where the energy flux is a positive constant,  right of the forcing scale. }
\label{fig:Flux}
 \end{center}
\end{figure}

Also for simulation {\bf A}, in Fig~\ref{fig:Flux} we plot the  averaged  non-dimensionalized flux $\tilde{\epsilon}_2(k)$ of the quadratic part of the energy, $\tilde{\mathcal{H}}_2$.
We observe a constant flux over a  region of $k$-space to the right of the forcing scale in agreement with the constant flux scale-invariant solutions of wave turbulence theory.  However, this flux falls off to a negative constant at high wave numbers, which we believe is down to the fact that numerically we can only measure the  flux of the {\em quadratic} energy, which represents the dominant part of the total energy only for weakly nonlinear waves. Thus, the deviation of  $\tilde{\epsilon}_2(k)$ from constant at large $k$ can be interpreted as an increase of nonlinearity at these scales. Note, however, that theoretically one predicts a decrease of nonlinearity at high $k$, see Eq.~\eqref{eq:ratioT}, which seems at odds with our numerical results - something that remains to be understood.

\begin{figure}
 \begin{center}
  \includegraphics[width=\columnwidth]{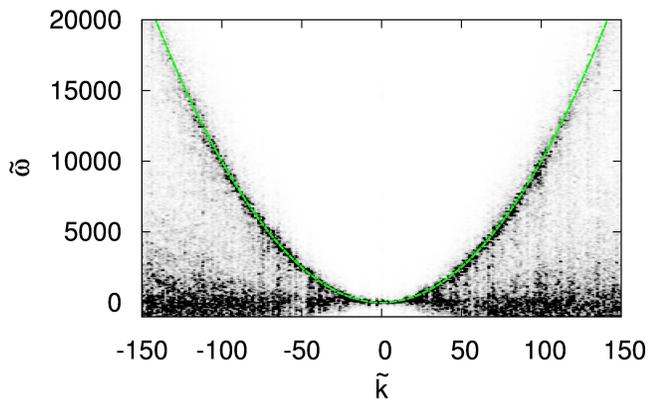}
\caption{The $({\bm k},\omega)$-plot of simulation {\bf A} representing   a normalized (separately for each ${\bm k}$) double Fourier transform of the wave amplitude $w(z,t)$.  The theoretical linear dispersion relation, $\tilde{\omega}  =  k^2$, is superimposed by the green dashed line.  }
\label{fig:kwplot-simA}
 \end{center}
\end{figure}

To add to this puzzle, we present a plot of the combined Fourier transform in both coordinate $x$ and time $t$ on the $(k, \omega)$-plane in
Figure~\ref{fig:kwplot-simA}. In addition to weakly nonlinear waves, which are narrowly distributed around the dispersion relation $\tilde{\omega} = k^2$, we observe a wide distribution centered at the zero frequency. The relative intensity of such a ``condensate" is higher at large $k$'s: its peak is about $1/3$ of the wave intensity at the lower end of the inertial interval $(k=50)$ and it becomes as strong as the wave intensity peak deeper in the
inertial range $(k=100)$. Again, we can offer no explanation for such a condensate (in the direct cascade range!) and to the fact why, in spite of the strong condensate, we perfectly observe the theoretical scaling of the Lvov-Nazarenko spectrum in the same range of wave numbers.

From the numerical data of Figs.~\ref{fig:Spectrum} and \ref{fig:Flux}, and with the addition of the numerical measurement of $\tilde{\Psi}$ in Eq.~\eqref{LN-spectrum} we can compute the numerical prefactor $C_{_{\rm LN}}$ of spectrum~\eqref{sol}, where  the formula for $C_{_{\rm LN}}$ expressed in terms of the non-dimensional parametrization is
\begin{equation}
C_{_{\rm LN}}=(4\pi)^{-1/3} \tilde{\epsilon}_k^{-1/3} \tilde{\Psi}^{2/3} \tilde{A}\ .
\end{equation}

The results from the simulations and the value of the numerical estimate of constant $C_{_{\rm LN}}$ are given in Table~\ref{tab:para}.

We observe remarkable agreement in the value of $C_{_{\rm LN}}$ in Eq.~\ref{sol} to the numerical data from simulation {\bf B}; see Table~\ref{tab:para}. Note that this is probably the first ever example of a wave turbulent system where the spectrum pre-factor obtained in numerics coincides with the one of the theoretical prediction. Indeed, realizing a pure wave turbulence state in numerical simulations is notoriously difficult because, on one hand, the waves must be weakly nonlinear and, on the other hand, they should not be too weak to overcome the finite size effects; see Sec.~\ref{ss:DNSgamma}. These effects become noticeable in our simulation {\bf A} and they become very clear in the simulation {\bf C} corresponding to the lowest forcing out of the three runs.

In simulation {\bf B}, with the forcing somewhat lower than in simulation {\bf A}, the spectrum and the flux are very similar to the ones shown in Figs.~\ref{fig:Spectrum} and \ref{fig:Flux}. The spectrum exhibits a scaling range with $-11/3$ exponent and is clearly different from the $-17/5$ prediction of Kozik and Svistunov. However, the value of the constant
$C_{_{\rm LN}}$ is slightly less than the theoretical one; see Table~\ref{tab:para}. This is probably due to the finite-size effects which manifest themselves in losses of wave resonances due to the discreteness of the ${\bm k}$-space
when the nonlinear frequency broadening $\Gamma_k$ becomes of the same order or less than the ${\bm k}$-grid spacing.

Such an effect is most prominent  in our weak-forcing simulation {\bf C} where the observed  constant $C_{_{\rm LN}}$ is an order of magnitude smaller; see Table~\ref{tab:para}.  In this run, we observed `sandpile' or `bursty' behaviour characterized by sudden spikes in the energies shown in Fig.~\ref{fig:Energy}.  We observe that all energies have become statistically stationary on the whole, however there is a clear bistable behaviour in the random changes to the energy that are intermittent in time.  In fact, the ratio of the two energies fluctuate between $|\tilde{\mathcal{H}}_6| / \tilde{\mathcal{H}}_2 \simeq 0.003 - 0.15$.

Obviously, the behaviour in simulation  {\bf C} is beyond the conditions which could be described by the standard kinetic equation. A qualitative description of this regime was presented in \cite{N06,meso,11Nazar} where the observed sandpile behaviour was originally conjectured. However, presently there is no rigorous theory describing this regime.

\subsection{Mesoscopic wave turbulence}

\begin{figure}
 \begin{center}
  \includegraphics[width=\columnwidth]{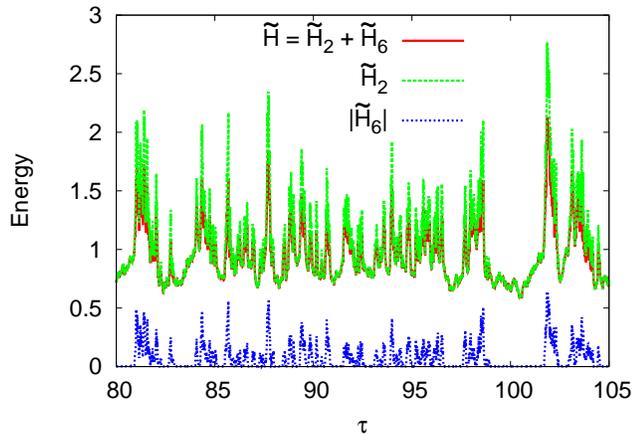}
\caption{Magnitudes of the linear, $\tilde{\mathcal{H}}_2$, nonlinear, $\tilde{\mathcal{H}}_6$ and total, $\tilde{\mathcal{H}}$ energies of simulation {\bf C} at late times in units of the linear timescale of the forcing mode: $\tau = 8\pi^2 /\kappa\Lambda k_f^2 $ where $k_f \approx 10$. We observe a statistically non-equilibrium stationary state of the energy. However, we also observe sharp jumps, peaks and troughs in the energy values.}
\label{fig:Energy}
 \end{center}
\end{figure}

\begin{figure}
 \begin{center}
  \includegraphics[width=\columnwidth]{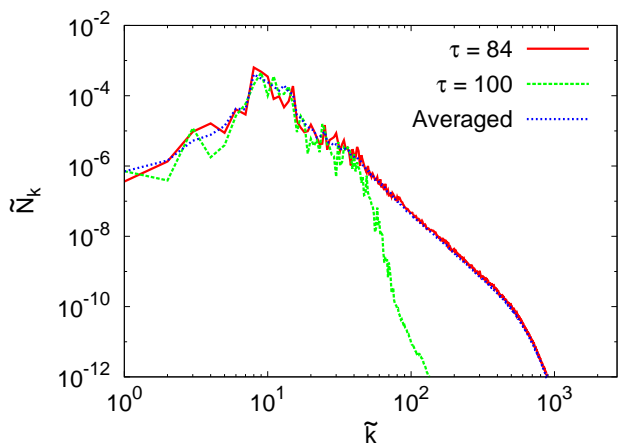}
\caption{Two wave action spectra at times $\tau = 84$ and $\tau = 100$, corresponding to a local maximum (red solid line) and a local minimum (green dashed line) in the energy respectively from simulation {\bf C}. Moreover, we compare these wave action spectra with that of the fully averaged wave action spectrum of Fig.~\ref{fig:Spectrum} (blue dotted line).}
\label{fig:fluct-spectra}
 \end{center}
\end{figure}

The sandpile behavior of simulation {\bf C} can also be seen in the evolution of the wave action spectrum $\tilde{N}_k$  presented in Fig.~\ref{fig:fluct-spectra} which shows two snapshots of the wave action spectrum at two times corresponding to a peak ($\tau = 84$) and trough ($\tau = 100$) of the energy.  In addition, in Fig.~\ref{fig:fluct-spectra} we include the fully averaged spectrum over the window $84 < \tau <100$ for comparison.  In  simulation $C$, we observe that the wave action spectrum $N_k$ oscillates between the two extreme states shown in Fig.~\ref{fig:fluct-spectra} at the same periods in which the energy fluctuations of peaks and troughs are seen
in Fig.~\ref{fig:Energy}. The spectrum fluctuates predominately in the high wave number region.

\begin{figure}
 \begin{center}
  \includegraphics[width=\columnwidth]{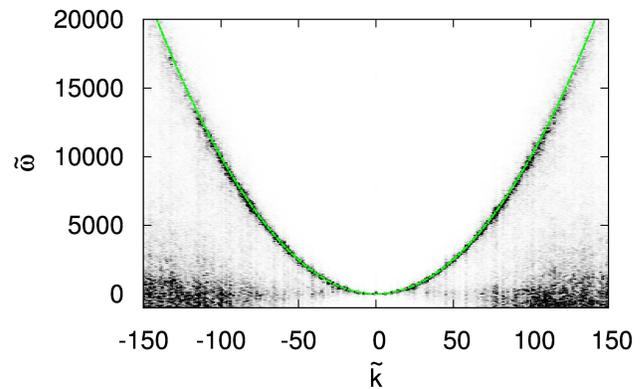}
\caption{The $({\bm k},\omega)$-plot of simulation {\bf C}. The random waves are characterized by the black curve, which is a normalized double Fourier transform of the wave amplitude $w(z,t)$.  The theoretical linear dispersion relation, $\omega_k = k^2$, is superimposed by the green dashed line.  }
\label{fig:kwplot-SimC}
 \end{center}
\end{figure}

In Fig.~\ref{fig:kwplot-SimC}, we present a numerical $({\bm k},\omega)$-plot for simulation {\bf C}, which is similar to the one discussed before for simulation {\bf A}, Fig.~\ref{fig:kwplot-simA}.
We see a qualitatively similar picture as before, but with a more narrow wave distribution and a weaker condensate around $\omega=0$ line, which is natural considering the lower level of nonlinearity in simulation {\bf C}.

\begin{figure}
 \begin{center}
  \includegraphics[width=\columnwidth]{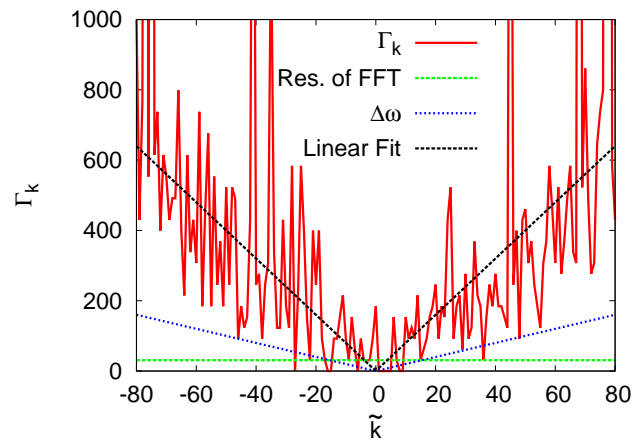}
\caption{We plot the nonlinear frequency broadening $\Gamma_k$ of the $({\B k},\omega)$-plot from simulation {\bf C} in the red solid line.  We overlay the resolution of the fast Fourier transform (green dashed line) and width of the frequency spacing $\Delta\omega$ (blue dotted line).  The black dashed line corresponds to a linear fit to $\Gamma_k$.}
\label{fig:kw-width}
 \end{center}
\end{figure}

Using Fig.~\ref{fig:kwplot-SimC} we can find the frequency resonance broadening $\Gamma_k$. This gives us a direct check if we are in a kinetic or mesoscopic wave turbulence regime \cite{N06,ZKPD05,meso,11Nazar}.
 Numerically, we compute $\Gamma_k$ by calculating the width of the interval along $\omega$ in which the dispersion relation in Fig.~\ref{fig:kwplot-SimC} is within half of maximum wave intensity at each wave number $k$.  We plot the result in Fig.~\ref{fig:kw-width}, along with the kinetic wave turbulence threshold $\Delta \omega$ and the limit of the resolution of the Fourier transform ($=2\pi / T$, where $T$ is the time window of the Fourier transform).  We observe in Fig.~\ref{fig:kw-width} that the frequency resonance broadening in the numerical simulation is larger than the resolution threshold, i.e. that the presented plot does indeed provide a useful information about $\Gamma_k$. Further, we see that
$\Gamma_k$ grows to higher $k$ consistently with theoretically predicted linear dependence \eqref{eq:gamma}.
Finally, we see that the values of $\Gamma_k$
are about factor three above the line corresponding to the threshold of mesoscopic wave turbulence $\Gamma_k \sim \Delta \omega$.  Since numerical factor of order unity are not controlled by such a simple estimate, we can conclude that such values of  $\Gamma_k$ are consistent with the condition of the mesoscopic wave turbulence regime.

Finally, it is also meaningful to ask if any  mesoscopic  effects, e.g.
the sandpile behaviour, can be seen in the physical space. We observe an interesting two-scale picture, which is consistent with the scale separation assumed in the non-local theory. Further, we observe that intensity of the slow (nearly horizontal) large-scale rolls suddenly drops at the moments corresponding to the drop of the total energy in Fig.~\ref{fig:Energy} marking onsets of the ``sandpile tip-overs''.
The same large-scale rolls are the main contributors into the low-frequency condensate component.

\begin{figure}
 \begin{center}
  \includegraphics[width=\columnwidth]{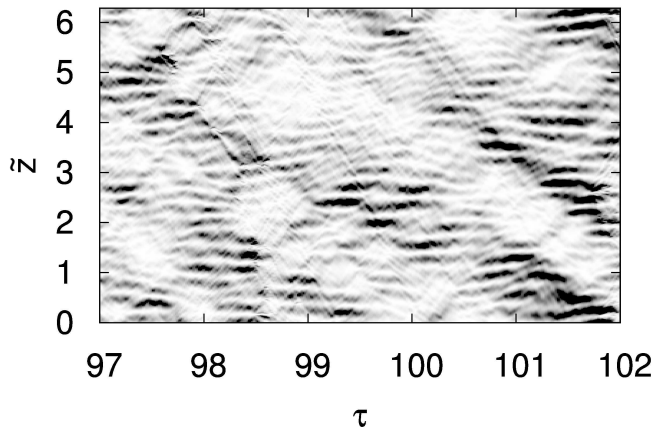}
\caption{We plot the physical space intensity $I(\tilde{z},\tilde{t}) = |w(\tilde{z},\tilde{t})|^2$ for simulation {\bf C} during a time period where the energy is fluctuating. Here black implies strong intensity and white implies low intensity.  The large scale structures observed correspond to the forcing scale.}
\label{fig:intensity}
 \end{center}
\end{figure}

\section{Conclusion}

In conclusion, we have shown that the spectrum proposed in~\cite{10LN} for Kelvin wave turbulence is the exact and unique solution of the kinetic equation and therefore is the one that should be physically realized.  All of the theoretical predictions have been confirmed via numerical simulations of the LNE, including both the scaling exponent and the spectrum pre-factor. We stress that the LNE is only applicable for nonlocal Kelvin wave turbulence. Thus, our numerical result that the (theoretically predicted) nonlocal spectrum is an attracting state demonstrates that the nonlocal dynamics is self-sustained and robust and the LNE model is self-consistent.

We also studied a mesoscopic regime of Kelvin wave turbulence which occurs at low amplitudes. We discovered that the system suffers from intermittent energy bursts that produce fluctuations in the wave action spectrum at high wave numbers.    This agrees with the sandpile scenario of evolution which was predicted for mesoscopic wave turbulence in
\cite{N06,ZKPD05,meso,11Nazar}.

Finally, numerically we observed some effects which we cannot presently explain. Namely, in addition to the weak dispersive waves we have found a strong low-frequency condensate component over a large range of scales covering the direct cascade range.
In the small-scale part of the inertial range the condensate is of a similar
strength to the wave component, and yet, somehow, it does not alter the theoretical (non-local) weak turbulence scaling.



\acknowledgments

We thank Reuven Zeitak for his help and encouraging discussions during this work.  We acknowledge the support of the U.S.-Israel Binational Scientific Foundation administrated by the Israeli Academy of Science, the Minerva Foundation,
Munich, Germany, through the Minerva Center for Nonlinear Physics at WIS and of the ANR program STATOCEAN (ANR-09-SYSC-014).

\end{document}